\documentclass[useAMS,usenatbib]{mn2e}

\usepackage{times}
\usepackage{graphicx}
\usepackage{subfigure}

\newcommand{\gsim}{\mathrel{\hbox{\rlap{\lower.55ex \hbox {$\sim$}}
                   \kern-.3em \raise.4ex \hbox{$>$}}}}
\newcommand{\lsim}{\mathrel{\hbox{\rlap{\lower.55ex \hbox {$\sim$}}
                   \kern-.3em \raise.4ex \hbox{$<$}}}}

\title[Radiative impact of stellar core formation]{Collapse of a molecular cloud core to stellar densities: the radiative impact of stellar core formation on the circumstellar disc}
\author[M.R. Bate]{Matthew R. Bate\thanks{E-mail:
mbate@astro.ex.ac.uk}\\ School of Physics, University of Exeter, Stocker
Road, Exeter EX4 4QL}

\date{\today}
\begin{document}
\maketitle
\begin{abstract}
We present results from the first three-dimensional radiation 
hydrodynamical calculations to follow the collapse of a molecular cloud 
core beyond the formation of the stellar core.  We find the energy 
released by the formation of the stellar core, within the 
optically-thick first hydrostatic core, is comparable to the binding 
energy of the disc-like first core.  This heats the inner regions of the 
disc, drives
a shock wave through the disc, dramatically decreases
the accretion rate on to the stellar core, and launches a temporary
bipolar outflow perpendicular to the rotation axis that travels in excess of
50 AU into the infalling envelope.

This outburst may assist the young protostar in launching
a conventional magnetic jet.  Furthermore, if these events are cyclic,
they may provide a mechanism for intense bursts of accretion separated
by long periods of relatively quiescent accretion which can potentially
solve both the protostellar luminosity problem and the apparent age 
spread of stars in young clusters.  Such outbursts
may also provide a formation mechanism for the chondrules found in meteorites, with the
outflow transporting them to large distances in the circumstellar
disc.
\end{abstract}
\begin{keywords}
accretion, accretion discs -- hydrodynamics -- radiative transfer -- stars: formation -- stars: low-mass, brown dwarfs -- stars: winds, outflows.
\end{keywords}

\section{Introduction}
\label{introduction}

More than four decades ago, \cite{Larson1969} performed 
the first numerical calculations of the collapse of a 
molecular cloud core to stellar core formation and beyond.  These one-dimensional
radiation hydrodynamical calculations revealed 
the main stages of protostar formation:
an almost isothermal collapse until the inner regions become
optically thick, the almost adiabatic formation of the first 
hydrostatic core (typical radius $\approx 5$~AU and initial mass
$\approx 5$~M$_{\rm J}$), the growth of this core as it accreted
from the infalling envelope, the second collapse within this
core triggered by the dissociation of molecular hydrogen,
the formation of the stellar core (initial radius $\approx 2$~R$_\odot$
and mass $\approx 1.5$~M$_{\rm J}$), and, lastly, the long
accretion phase of the stellar core to its final mass.
Subsequent one-dimensional \citep[e.g.][]{MasInu2000}
and two-dimensional \citep{Tscharnuter1987, Tscharnuteretal2009} calculations 
have not changed this qualitative picture substantially,
although the latter have allowed
the disc-like structure of rotating first cores to be studied.

The first three-dimensional hydrodynamical calculations 
to follow the collapse to stellar core formation 
were performed more than a decade ago by \cite{Bate1998}.  
These calculations of rotating molecular cloud cores
examined the non-axisymmetric evolution of the first core 
and the second collapse phase.  If the first core was 
rotating rapidly enough it was found to be dynamically
unstable to a bar-mode leading to the formation of
trailing spiral arms.  Gravitational torques removed 
angular momentum and rotational support from the 
inner regions of the first core, quickening the onset
of the second collapse and preventing fragmentation
during the second collapse phase to form close 
binaries.  Several subsequent studies have 
investigated this phenomenon in more detail 
\citep*{SaiTom2006, SaiTomMat2008, MacInuMat2010},
some also including magnetic fields and finding outflows
\citep{Machidaetal2005, MacInuMat2006}.  However, all these
calculations used 
barotropic equations of state rather than solve the
radiation hydrodynamical problem.

\begin{figure*}
\centering \vspace{-2.7cm}
    \includegraphics[width=15.5cm]{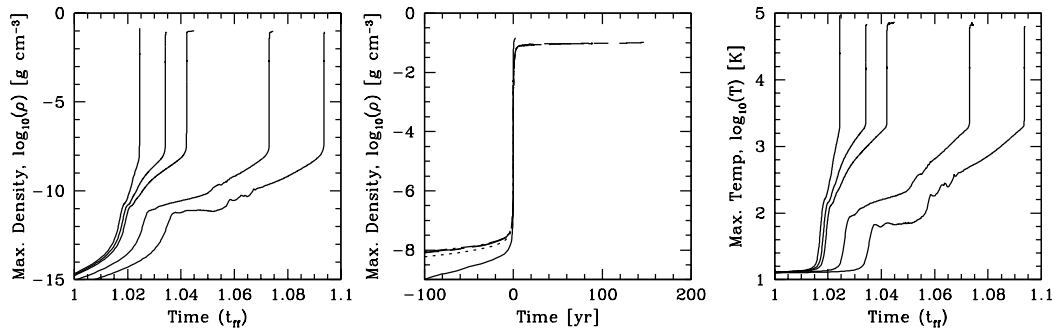}\vspace{-13.3cm}
\caption{The time evolution of the maximum (central) density (left two panels) and temperature (right panel).  For the left and right panels, the lines are for cloud cores with $\beta=0,5\times 10^{-4},0.001,0.005,0.01$ from left to right.  For the central panel, for each calculation the time in years has been zeroed when the density first exceeded $10^{-3}$~g~cm$^{-3}$ and the $\beta=0$ calculation is plotted with a solid line, $\beta=5\times 10^{-4}$ with a dotted line, and the other calculations are indistinguishable. The free-fall time of the initial cloud core, $t_{\rm ff}=1.8\times 10^{12}$~s (56,500 yrs).  
Each calculation was performed with $10^6$ SPH particles.
}
\label{evolutions}
\end{figure*}

The first three-dimensional calculations including radiative transfer that followed 
collapse to the point of stellar core formation (but not beyond)  were \citet{WhiBat2006}, using the 
flux-limited diffusion approximation, and \cite{Stamatellosetal2007},
using a radiative cooling approximation.
Most recently, radiation magnetohydrodynamical calculations
of cloud collapse have been performed \citep{Tomidaetal2010}, but were stopped 
before the onset of the second collapse.


In this paper, we report results from the first three-dimensional 
radiation hydrodynamical calculations to follow the 
collapse of rotating molecular cloud cores {\it beyond} the formation
of the stellar core.  We find that the use of 
radiation hydrodynamics rather than a barotropic
equation of state has little effect up until the formation
of the stellar core.  However, with radiative transfer, the energy 
released by the formation of the stellar core has a
dramatic effect on the surrounding disc and envelope
and launches a temporary outflow {\it even in the absence of a 
magnetic field}.

\section{Computational method}
\label{method}

The calculations presented here were performed 
using a three-dimensional smoothed particle
hydrodynamics (SPH) code based on the original 
version of \citeauthor{Benz1990} 
(\citeyear{Benz1990}; \citealt{Benzetal1990}), but substantially
modified as described in \citet{BatBonPri1995},
\citet*{WhiBatMon2005}, \citet{WhiBat2006},
\cite{PriBat2007}, and 
parallelised using both OpenMP and MPI.

Gravitational forces between particles and a particle's 
nearest neighbours are calculated using a binary tree.  
The smoothing lengths of particles are variable in 
time and space, set iteratively such that the smoothing
length of each particle 
$h = 1.2 (m/\rho)^{1/3}$ where $m$ and $\rho$ are the 
SPH particle's mass and density, respectively
\cite[see ][for further details]{PriMon2007}.  The SPH equations are 
integrated using a second-order Runge-Kutta-Fehlberg 
integrator with individual time steps for each particle
\citep{BatBonPri1995}.
To reduce numerical shear viscosity, we use the
\cite{MorMon1997} artificial viscosity
with $\alpha_{\rm_v}$ varying between 0.1 and 1 while $\beta_{\rm v}=2 \alpha_{\rm v}$
\citep[see also][]{PriMon2005}.

\subsection{Equation of state and radiative transfer}

We use an ideal gas equation of state 
$p= \rho T \cal{R}/\mu$, where 
$T$ is the gas temperature, $\cal{R}$ is the gas constant, 
and $\mu$ is the mean molecular
mass.  The equation of state takes into account the translational,
rotational, and vibrational degrees of freedom of molecular hydrogen 
(assuming a 3:1 mix of ortho- and para-hydrogen; see
\citealt{Boleyetal2007}).  It also includes molecular
hydrogen dissociation, and the ionisations of hydrogen and helium.  
The hydrogen and helium mass fractions are $X=0.70$ and 
$Y=0.28$, respectively.
The contribution of metals to the equation of state is neglected.
Two temperature (gas and radiation) radiative transfer in the flux-limited
diffusion approximation is implemented as described by \citet{WhiBatMon2005}
and \citet{WhiBat2006}, except that the standard explicit SPH contributions to the gas energy equation due to the work and artificial viscosity are used when solving the (semi-)implicit energy equations to provide better energy conservation.  We assume solar metallicity gas, using 
the interstellar grain opacity tables of \citet{PolMcKChr1985} and the gas opacity
tables of \citet{Alexander1975} (the IVa King model)  \citep[see][]{WhiBat2006}.

\subsection{Initial conditions}
\label{initialconditions}

The initial conditions for the calculations are identical to those 
of \cite{Bate1998}.  We follow the collapse 
of initially uniform-density, uniform-rotating, molecular cloud 
cores of mass $M=1 {\rm \ M}_\odot$ and
radius $R=7 \times 10^{16} {\rm \ cm}$.  The ratios of the thermal
and rotational energies to the magnitude of the gravitational 
potential energy are $\alpha = 0.54$ and $\beta$, respectively.  We have performed
calculations with $\beta=0$ (not rotating),
and $\beta = 5\times 10^{-4}, 0.001, 0.005$ and 0.01.

To satisfy the resolution criterion of \cite{BatBur1997} that the 
minimum Jeans mass during the calculation contains at least 
$\approx 2 N_{\rm neigh} = 100$ particles, we require at 
least $1 \times 10^5$ equal-mass particles.  To test for convergence, 
we performed calculations using $1 \times 10^5$, $3 \times 10^5$, $1 \times 10^6$,
and $3\times 10^6$ equal-mass SPH particles.
The calculations were performed on the University of Exeter 
Supercomputer, an SGI Altix ICE 8200.

\begin{figure*}
\centering \vspace{-0.0cm}
    \includegraphics[width=17.5cm]{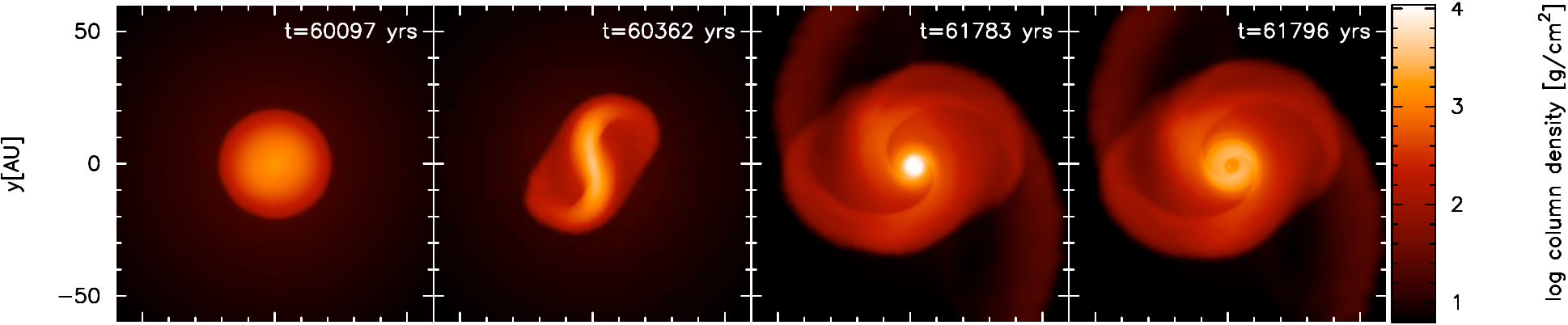}
    \includegraphics[width=17.5cm]{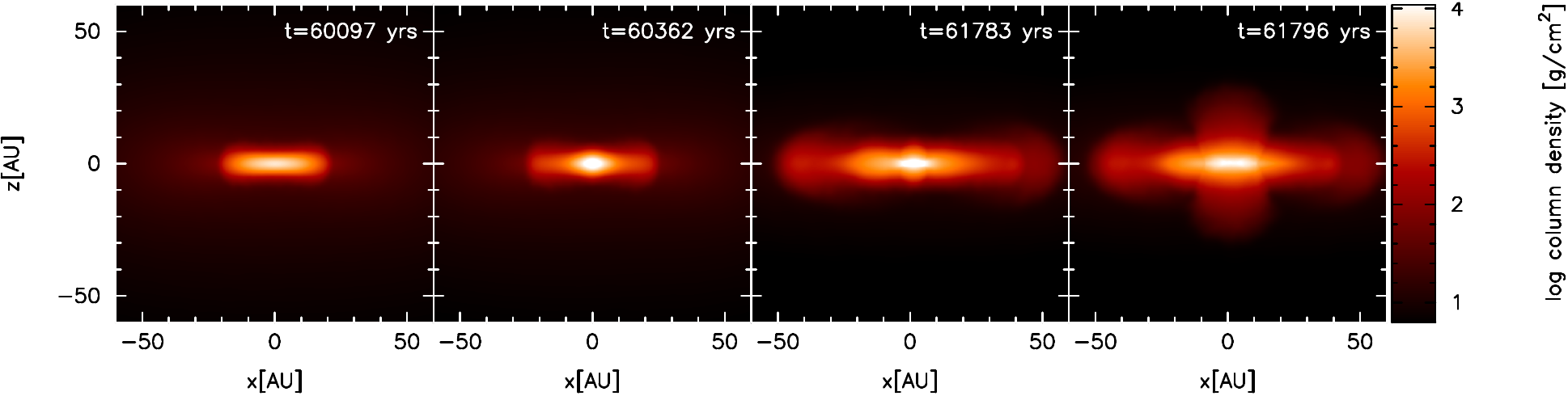}
\caption{Column density images of the evolution of the first core, shock wave, and outflow for the $\beta=0.005$ case (using $3\times 10^6$ SPH particles).  The rapidly rotating first core is highly oblate (left panels) and undergoes a dynamical bar-mode instability (centre-left panels) which transports angular momentum out of the centre, transforms the core into a disc, and triggers the formation of the stellar core.  The energy released by stellar core formation heats the inner regions of the disc and drives a shock wave through the disc (upper right two panels) and a bipolar outflow perpendicular to the disc (lower right two panels). Animations can be found at http://www.astro.ex.ac.uk/people/mbate/Animations/Stellar/ .}
\label{images}
\end{figure*}

\section{Results}
\label{results}

The collapse of the molecular cloud cores up until the formation of the stellar
core proceeds in a manner very similar to that reported from previous 
three-dimensional calculations using barotropic equations of state
\citep{Bate1998, SaiTom2006, SaiTomMat2008, MacInuMat2010}.  Fig.~\ref{evolutions}, gives the evolution of the maximum density 
and temperature for the
$10^6$ particle calculations with different initial rotation rates.
The initial collapse is almost isothermal
until the maximum density exceeds $\approx 10^{-13}$~g~cm$^{-3}$.
As the central regions of the cloud become optically thick, the gas begins
to trap the radiation and the collapse enters an almost adiabatic phase
where the temperature rises as the gas is compressed.  This leads to the
formation of a pressure-supported `first core' \citep{Larson1969}, which can be seen in Fig.~\ref{evolutions} when the initial collapse
stalls with central (maximum) densities $\sim 10^{-11}$~g~cm$^{-3}$ and 
temperatures of $\approx 70-120$~K, depending on the degree of
rotational support (i.e. cores that rotate more quickly have lower
central temperatures).  Without
rotation, the first core has an initial mass of $\approx 5$ Jupiter masses (M$_{\rm J}$)
and a radius of $\approx 5$~AU.  However, with higher initial rotation rates 
of the molecular cloud core,
the first cores become progressively more oblate.  For example, with $\beta = 0.005$ before the onset of dynamical instability, the first core has a radius of $\approx 20$~AU and a major to minor axis ratio of $\approx$4:1 (left panels, Fig.~\ref{images}).  Thus,
for the higher rotation rates, the first core is actually a disc, but without a central
object.  As pointed out by \cite{Bate1998} and \cite{MacInuMat2010}, 
in these cases the disc actually forms {\it before} the star.

The subsequent evolution of the first core depends on its rotation rate.
Non-rotating and slowly rotating cores evolve as they accrete mass from the
surrounding infalling envelope with their central densities and temperatures increasing
(Fig.~\ref{evolutions}, calculations with $\beta \le 0.001$).  When the
central temperatures of a first core exceeds $\approx 2000$~K, molecular 
hydrogen begins to dissociate, absorbing energy and leading to a second 
hydrodynamic collapse deep within the first core.  The collapse 
continues until the molecular hydrogen has been complete dissociated at
the centre of the core where upon a 
second pressure-supported core begins to form,
the `stellar' core \citep{Larson1969}.  The formation of the stellar core occurs just
a few years after the onset of the second collapse, during which the maximum density 
increases from $\sim 10^{-8}$ to $\approx 0.1$ g~cm$^{-3}$ and the maximum temperature
increases from 2000 to $>60,000$~K.  The stellar core is formed with a mass
of $M_{\rm sc} \approx 1.5$~M$_{\rm J}$ and a radius of $R_{\rm sc} \approx 2$~R$_\odot$.  Without
rotation, the stellar core accretes the remnant of the first core in which it is embedded
in $\approx 10$~yr and then accretes the envelope \citep{Larson1969}.

If the first core is rotating rapidly enough that its value of $\beta > 0.274$, the
core is dynamically unstable to the growth of non-axisymmetric structure \citep{Bate1998}.
For the particular initial conditions used here, this occurs for the $\beta=0.005$ (see Fig.~\ref{images})
and $\beta=0.01$ cases.  The first core develops a bar-mode, and then
spiral structure as the ends of the bar wind up \citep[e.g.][]{Durisenetal1986,Bate1998}.
The spiral structure removes angular momentum from the inner parts of the first
core, the effect of which can be seen in the evolution of density and temperature in
Fig.~\ref{evolutions}.  Specifically, for the $\beta=0.005$ case, the slow increase in
central density and temperature accelerate with the onset of the spiral structure at
$t=1.05~t_{\rm ff}$.  This substantially accelerates the evolution of the first core towards 
the second collapse, which would have taken much longer to reach without the
angular momentum redistribution.  It also inhibits fragmentation during the second
collapse phase because the gas undergoing the second collapse has less angular
momentum than it would otherwise \citep{Bate1998}.  When the stellar core forms
it is {\it already surrounded by a large disc} (radius $\approx 50$~AU; Fig.~\ref{images}) which is the remnant of
the gravitationally-unstable first core \citep{Bate1998, MacInuMat2010}.

For even higher rotation rates of the initial molecular cloud core
($\beta \gsim 0.02$), the first core is actually a ring-like structure \citep[e.g.][]{NorWil1978,ChaWhi2003} which is 
prone to dynamical fragmentation into two or more objects, each of which evolves in a
similar manner to the more slowly rotating first cores.

As mentioned at the start of this section, all of this evolution with radiation hydrodynamics
is almost identical to the evolution that is obtained when using a barotropic equation of
state.  The small differences that are found will be discussed in a separate paper.  The main purpose of this letter is to discuss the evolution subsequent to the formation of the
stellar core, which is {\it qualitatively different} to that obtained with a barotropic equation
of state.

\subsection{The effect of stellar core formation}

When using a barotropic equation of state, the formation of the stellar core deep within
the optically-thick disc has no effect on the temperature of the majority of the
gas in the disc because the temperature is set purely according to the density of the gas.
However, with radiation hydrodynamics, the situation is completely different.

\begin{figure}
\centering \vspace{-0.0cm}
    \includegraphics[width=7.5cm]{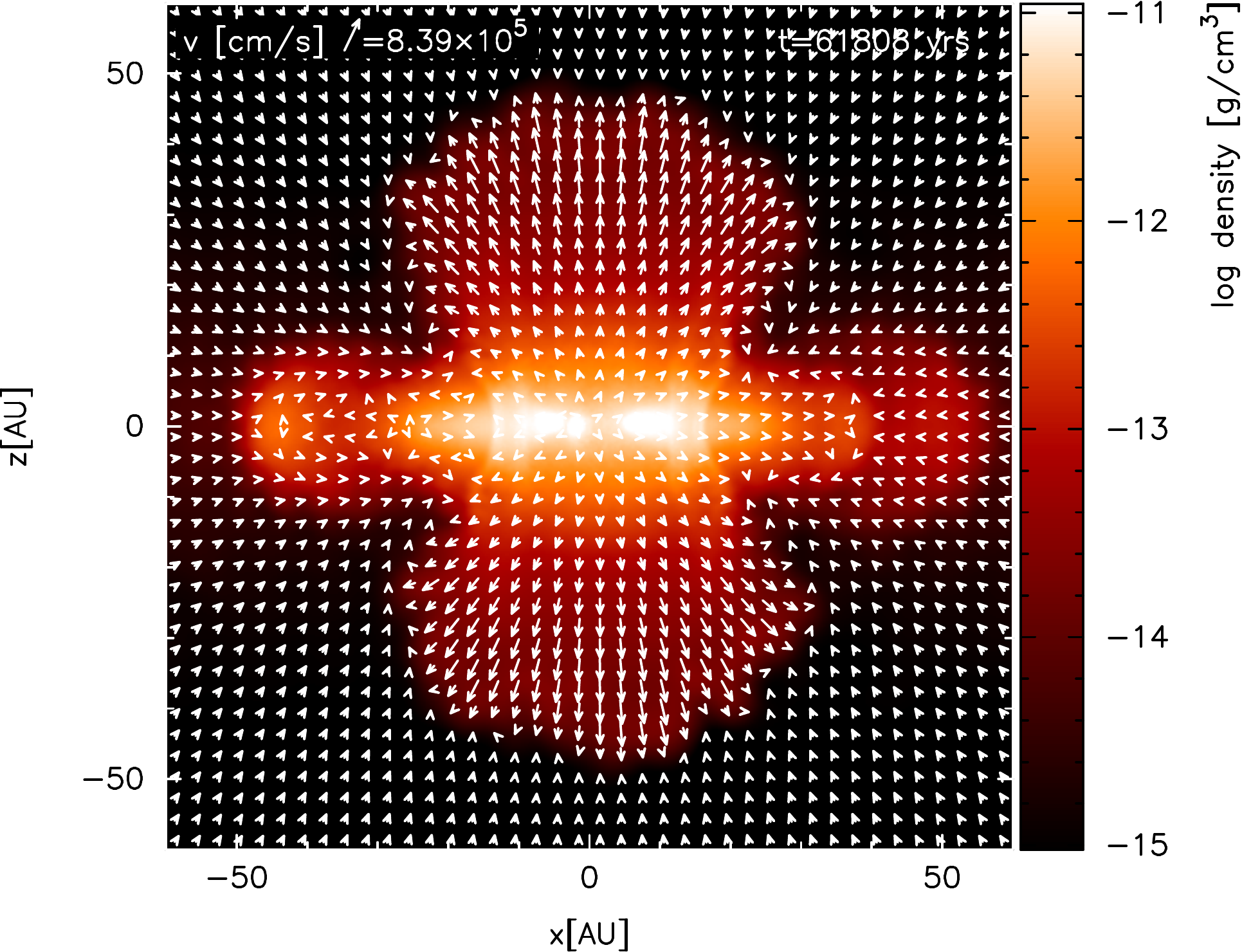}
\caption{A cross section down the rotation axis ($x-z$ plane) of the $\beta=0.005$ calculation with $3\times 10^6$ particles showing the density (colour scale) and velocities (vectors).  The bipolar outflow is clearly visible and has travelled 50~AU into the surrounding infalling envelope only 35 yr after stellar core formation.  The shock wave travelling along the midplane of the disc has propagated 17~AU out into the disc in the same period of time. Animations can be found at http://www.astro.ex.ac.uk/people/mbate/Animations/Stellar/ .}
\label{vectors}
\end{figure}

\begin{figure}
\centering \vspace{-0.0cm}
    \includegraphics[width=7.0cm]{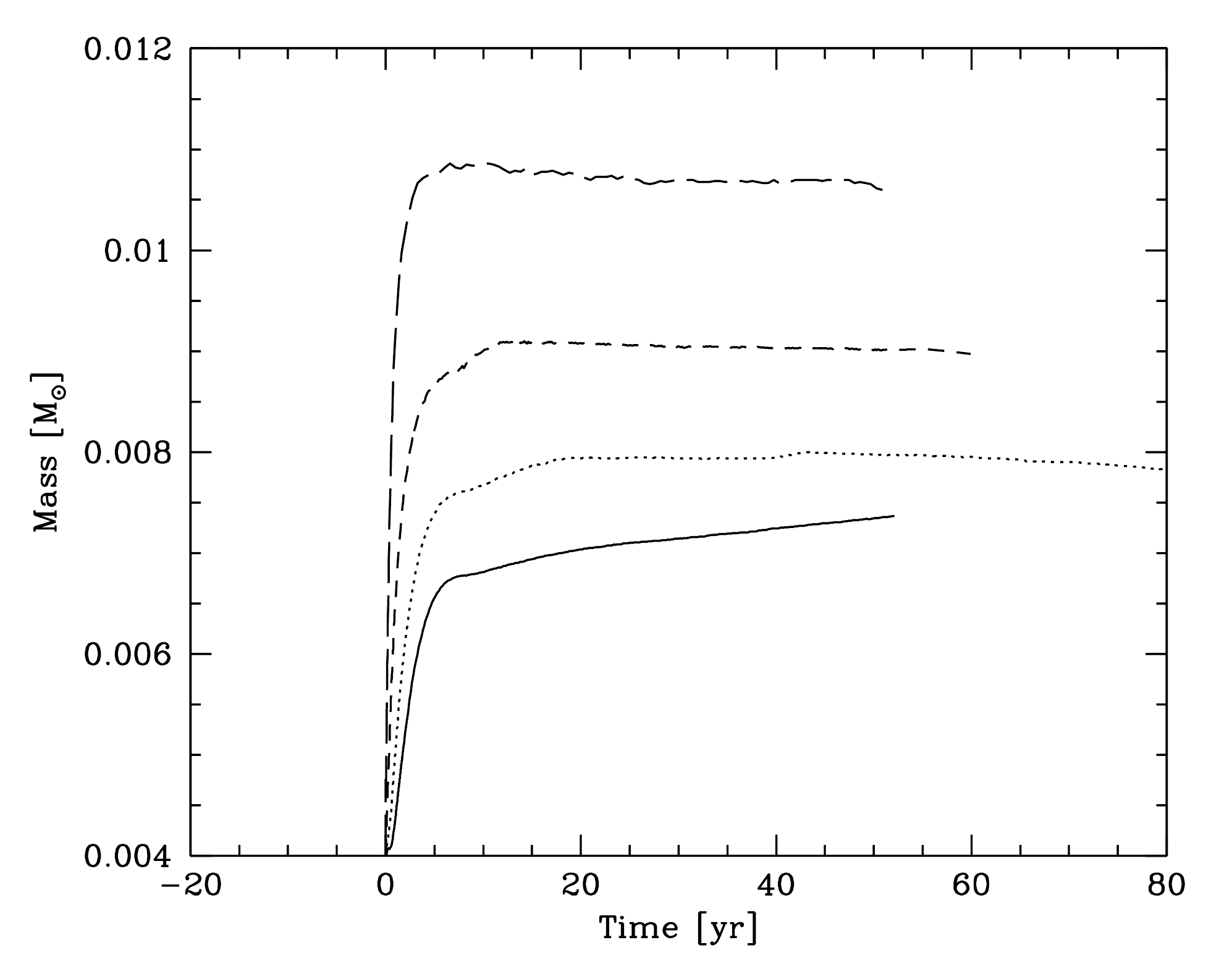}
\caption{The mass of the stellar core (gas with density $>10^{-4}$~g~cm$^{-3}$) measured from the time of its formation for the $\beta=0.005$ molecular cloud core performed using $10^5$ (long-dashed line), $3\times 10^5$ (short-dashed line), $10^6$ (dotted line), and $3\times 10^6$ (solid line) particles.  With low resolution the radiative feedback stops accretion onto the stellar core, but for the highest resolution the accretion continues at a low level ($10^{-5}$~M$_\odot$~yr$^{-1}$). }
\label{convergence}
\end{figure}

When the second collapse occurs and produces the stellar core, the gravitational
potential energy that is released is $\sim GM_{\rm sc}^2/R_{\rm sc} = 4 \times 10^{42}$~erg.
Since the stellar core is in virial equilibrium, approximately half of this energy is
to be radiated away.  Moreover, the stellar core rapidly begins to accrete from the
first core, reaching a mass of $\approx 6$~M$_{\rm J}$ in only a few years (i.e. more 
quickly than the dynamical timescale of the disc; see below).  This increases the total 
energy released to $\approx 3 \times 10^{43}$~erg.
However, the binding energy of the disc in the $\beta=0.005$ calculation, for example, 
calculated just before the onset of the second collapse (i.e. the binding energy of the 
gas surrounding the stellar core that makes up the disc) is only $4 \times 10^{43}$~erg 
(which can be estimated as $\sim GM_{\rm d}^2/R_{\rm d}$ with  
$M_{\rm d} \approx 0.18~{\rm M}_\odot$ and taking a mean `spherical' radius of $R_{\rm d} \approx 15$~AU).
Thus, the disc suddenly finds itself irradiated from the inside by an energy source 
emitting a substantial fraction of the binding energy of the disc itself.
Because the disc is extremely optically thick, this energy is temporarily trapped in the centre of the disc
and heats the gas dramatically, sending a weak shock wave (Mach number $\approx 1$) out along
the midplane of the disc (initial speed $\approx 2$~km~s$^{-1}$).  However, perpendicular to the disc, the effect is even
more dramatic.  Because there is less material along the rotation axis, the hot gas
finds it easiest to break out in this direction and a bipolar outflow is launched.
Whereas the wave within the disc decays as it travels leaving the
bulk of the disc gravitationally bound, the gas forming the bipolar outflow
has velocities in excess of 8 km~s$^{-1}$ and travels out into the infalling 
envelope to distances in excess of 50~AU in less than 50 years 
(Figs.~\ref{images} and \ref{vectors}).

The energy released by stellar core formation is somewhat dependent
on the resolution of the calculations because the heating and outflow inhibit 
accretion onto the stellar core, which in turn results in less energy being released.
Thus, the higher the resolution, the quicker the radiative feedback is able to inhibit
the accretion and, therefore, the lower the initial mass of the stellar core.
In Fig.~\ref{convergence}, we plot the mass with density $>10^{-4}$~g~cm$^{-3}$
versus time for the $\beta=0.005$ case with resolutions ranging from $10^5$ to
$3\times 10^6$ SPH particles.  It can be seen that the mass of the stellar core at which the
feedback dramatically curtails the accretion 
decreases from $\approx 11$ to 6~M$_{\rm J}$ with increased resolution.
For a few years, the stellar core grows at a rate of $\sim 10^{-3}$~M$_\odot$~yr$^{-1}$
(even with the highest resolution).  
With low-resolution ($\le 1\times 10^6$ SPH particles),
the accretion onto the stellar core actually ceases entirely during the launching of the
outflow.  However, using $3\times 10^6$
SPH particles, we find that the stellar core does continue to grow during this time, but at
a much reduced accretion rate of $1\times 10^{-5}$~M$_\odot$~yr$^{-1}$ (measured between 
20 and 50 years after stellar core formation; Fig.~\ref{convergence}).
Although convergence with increasing resolution is relatively slow, a strong outflow is
launched in all cases.

\section{Discussion}

These first calculations raise many questions to be
followed up in the future.  First, although 
the outflow seen in the calculations discussed here is transient and cannot be 
the origin of protostellar jets or outflows, such heating upon stellar core formation 
might aid in the launching of a magnetic jet by helping to clear 
material perpendicular to the disc and combining substantial thermal support with
magnetic support.  Therefore, using future
radiation magnetohydrodynamical calculations it will be important to assess the
relative roles of the radiative impact of stellar core formation and magnetic fields in the
launching of the protostellar jet.  Similarly, it would be desirable to resolve the accretion 
shock at the surface of the stellar core and to use a radiative transfer method that treats the 
optically-thin regime better than the flux-limited diffusion approximation.

Second, although the dramatic decrease in the stellar core
accretion rate is also a transient, 
the question arises as to whether such a situation might be cyclic.  One
can imagine that after the disc settles down again, the accretion rate on to
the stellar core may increase, leading to a repeat of the rapid accretion, the generation 
of a large accretion luminosity, the strengthening of an outflow and heating of the disc,
and the consequential decrease in the stellar core's accretion rate.  The driving
mechanism for the increased stellar accretion rate could be the redevelopment of
spiral density waves in the disc as it recovers from the previous shock wave (Fig.~\ref{images}).
Such gravitational instabilities have been invoked to argue that accretion in the
early Class 0 and I stages of star formation may be highly variable 
\citep{VorBas2005, VorBas2006}.  The stellar accretion luminosity may then provide
a way to temporarily switch the gravitational instability off.
Another possible mechanism for variable accretion might be a mismatch between
the accretion rates given by gravitational instability at large radii and the magnetorotational
instability at small radii \citep{ArmLivPri2001,ZhuHarGam2009}.
If most of a star's mass is accumulated in short bursts of intense accretion,
this can potentially explain the observation that protostars are observed to be less luminous than
expected for steady accretion \citep[see][]{Kenyonetal1990, HarKen1996}, and also 
the apparent age spread of young stars in the
Hertzsprung-Russell diagram \citep*[see][]{BarChaGal2009}.

Finally, there is the age old problem of how the chondrules
found in meteorites were formed \citep{Grossmanetal1988} as their production is thought to
require high temperatures ($>1500$~K).  Although
many theories have been advanced in order to explain chondrule formation
\citep[see, for example, the review by][]{Boss1996}, there
is still no widely accepted mechanism.  In this context, it is of interest to note that
the temperatures within $\approx 3$~AU of the stellar core 
at the launching of the outflows exceed 1500~K.  While most of this material
is no doubt be accreted by the star, a fraction produces the outflows 
which reach beyond $50$~AU.  If chondrules were formed
in this outburst event, some might be entrained in the outflow and eventually
fall back into the circumstellar disc at large distances from the star.  Again, if
such events were cyclic this would help because chondrule formation is thought
to have occurred over several million years \citep{Scott2007}.

\section{Conclusions}
\label{conclusions}

We have presented results from the first three-dimensional radiation 
hydrodynamical calculations to follow the collapse of a molecular 
cloud core beyond the formation of the stellar core.  We find the evolution 
before the formation of the stellar core is very similar to that found 
in the past using barotropic equations of state.  In particular,
the evolution of the first hydrostatic core which, depending
on its rotation rate, may be dynamically unstable to the growth
of non-axisymmetric perturbations and the generation of spiral structure,
is very similar to that found in barotropic calculations.  As found in
earlier calculations, a rapidly rotating first core actually evolves into a disc so that
the disc actually forms {\it before} the stellar core.

However, we find that the evolution following the formation of the stellar
core is {\it qualitatively} different in radiation hydrodynamical calculations.
In barotropic calculations, the formation of the stellar core deep
inside the first core (or disc) has no effect on the surrounding disc because
the temperature of the gas is simply set by the density of the gas.
However, with radiation hydrodynamics, the energy released by the formation
of the stellar core within the optically thick disc is similar to the binding 
energy of the disc.  This heats the inner regions of the disc, drives
a shock wave outwards through the disc, dramatically decreases
the accretion rate on to the stellar core, and launches a
bipolar outflow perpendicular to the rotation axis that can travel in excess of
50 AU out into the infalling envelope in less than 50 years.

We speculate that such outflows may assist the young protostar in launching
a conventional magnetic jet by clearing a path perpendicular to the disc
and adding substantial thermal pressure to the force provided by the magnetic field.  It may also be that such 
events are cyclic, occurring every time the accretion rate onto the protostar
exceeds a certain level rather than simply being a one-off event associated 
with the formation of the stellar core.  If so, such events may 
provide the mechanism for intense bursts of accretion separated
by long periods of relatively quiescent accretion which may be necessary to solve 
the protostellar luminosity problem and the apparent age spread of young
stars.   Finally, such outbursts
may provide another mechanism for the formation of chondrules in meteorites and the associated 
outflows may be able to transport them to large distances in the circumstellar
disc.

\section*{Acknowledgments}

The computations were performed using the University of Exeter Supercomputer. 
Figs. \ref{images} and \ref{vectors} were produced using the publicly available SPLASH visualisation software \citep{Price2007}.
This work, conducted as part of the award ``The formation of stars and planets: Radiation hydrodynamical and magnetohydrodynamical simulations" made under the European Heads of Research Councils and European Science Foundation EURYI (European Young Investigator) Awards scheme, was supported by funds from the Participating Organisations of EURYI and the EC Sixth Framework Programme. 

\bibliography{mbate}

\end{document}